\documentclass[manuscript,onecolumn]{emulateapj}

\newcommand{\be}{\begin{equation}}
\newcommand{\ee}{\end{equation}}

\slugcomment{accepted to AJ}
\shorttitle{Adaptive Scheduling}
\shortauthors{Ford}
\begin{document}

\title{Adaptive Scheduling Algorithms for Planet Searches}

\author{Eric B.\ Ford\altaffilmark{1,2,3,4,5}}

\altaffiltext{1}{Department of Astrophysical Sciences, 
	Princeton University, 
	Peyton Hall, 
	Princeton, NJ 08544-1001, USA}
\altaffiltext{2}{Astronomy Department, 
	601 Campbell Hall, 
	University of California at Berkeley, 
	Berkeley, CA 94720-3411, USA}
\altaffiltext{3}{Harvard-Smithsonian Center for Astrophysics,
        MS-51,
	60 Garden Street,
	Cambridge, MA 02138, USA}
\altaffiltext{4}{Hubble Fellow}
\altaffiltext{5}{present address: Department of Astronomy,
        University of Florida,
	211 Bryant Space Science Center,
	P.O. Box 112055, 
	Gainesville, FL, 32611-2055, USA}

\begin{abstract}
High-precision radial velocity planet searches have surveyed
over $\sim\!2000$ nearby stars and detected over $\sim\!200$ planets.  While
these same stars likely harbor many additional planets, they will
become increasingly challenging to detect, as they tend to have
relatively small masses and/or relatively long orbital periods.
Therefore, observers are increasing the precision of their
observations, continuing to monitor stars over decade timescales, and
also preparing to survey thousands more stars.  Given the considerable
amounts of telescope time required for such observing programs, it is
important use the available resources as efficiently as possible.
Previous studies have found that a wide range of predetermined
scheduling algorithms result in planet searches with similar
sensitivities.  We have developed adaptive scheduling algorithms which
have a solid basis in Bayesian inference and information theory and
also are computationally feasible for modern planet searches.  We have
performed Monte Carlo simulations of plausible planet searches to test
the power of adaptive scheduling algorithms.  Our simulations
demonstrate that planet searches performed with adaptive scheduling
algorithms can simultaneously detect more planets, detect less massive
planets, and measure orbital parameters more accurately than
comparable surveys using a non-adaptive scheduling algorithm.  We
expect that these techniques will be particularly valuable for the
N2K radial velocity planet search for short-period planets as well as
future astrometric planet searches with the Space Interferometry
Mission which aim to detect terrestrial mass planets.
\end{abstract}


\section{Introduction}

Radial velocity planet searches have surveyed over 2000 nearby solar
type stars and discovered over 200 planets.  The surveys require many
high precision radial velocity observations of each star in the
survey, and hence a significant amount of observing time.  
For example, the N2K project has recently begun surveying the next
$\sim 2000$ nearby stars for planets.  This project aims to discover
dozens of hot Jupiters and hopefully additional transiting planets
(Fischer et al.\ 2004).  Given the large target list for the N2K
project, it is essential that inferences about the presence of planets
and their orbits be made as efficiently as possible.  The
observing program aims to take three observations of each star on
consecutive nights, followed by an additional observation one to a few
months later.  In many cases, it is clear after a few observations
that the radial velocity observations have a dispersion significantly
greater than would be expected due to measurement errors only.
However, there may still be a large range of possible orbital
solutions, and several additional observations will often be
required to determine the planetary orbits.  Given the considerable
observation time required for such planet searches and the value of
telescope time, it is important that these surveys be as
efficient as possible.  

Previous studies have demonstrated that a variety of observing
schedules result in comparable efficiencies for detecting planets
(Sozzetti 2002, Ford 2004).  However, these studies have only
considered {\em non-adaptive } observing schedules, i.e., schedules
that are fully determined before any observations are taken.  In this
paper, we develop and test the efficiency of {\em adaptive}
scheduling algorithms.  We describe how adaptive scheduling algorithms
can help increase the efficiency of planet searches by utilizing
the information available from the previous observations to plan future
observations.


Adaptive scheduling algorithms can be based on Bayesian
inference (Loredo 2004).  Within this framework, the need to make decisions is
minimized.  For example, the experimenter does not say that a planet
has been detected, but rather states that the posterior probability for
the null hypothesis is some small value.  This eliminates the need to
chose threshold functions.  More importantly, by eliminating
decisions and basing the scheduling algorithm on the posterior
probability distribution, the posterior probability distribution is
not biased by the choice of observing schedule.  Additionally, it is
straightforward to test new hypotheses which will inevitably be
formulated after making some observations.  For these reasons, we have
developed adaptive scheduling algorithms within the Bayesian
framework, always considering all possible models (weighted according
to their posterior probability).  

It should be noted that the use of adaptive scheduling algorithms ---
even those based on Bayesian inference--- does affect the distribution
of the posterior distributions.  Indeed, that is the purpose of
employing such algorithms.  For example, if an adaptive scheduling
algorithm is chosen to increase the sensitivity of an observing
program for detecting planets, then it is expected that an ensemble of
surveys employing the adaptive scheduling algorithm will result in
detecting more planets than a comparable ensemble of surveys using a
fixed scheduling algorithm.  It should be noted that this is also true
of non-adaptive scheduling algorithms.  For example, an observing
program which makes observations over a long duration will be
sensitive to planets with orbital periods comparable or less than the
survey duration, but there may be ambiguities in determining the
orbital period associated with aliasing.  If an observing program
obtained the same number of observations with the same precision, but
performed all the observations during a smaller interval of time, then
the shorter survey would be less sensitive to planets with orbital
periods longer that the short survey's duration, but would likely
reduce aliasing ambiguities for short period planets.  Thus, when
analyzing the properties of a population of stars and planets, it is
alway important to account for the scheduling algorithm used.

Our approach applies the principles of Bayesian inference and
information theory to guide the choice of observation times (and later
choice of target stars).  Following Loredo (2004), we assume a prior
for both the probability that each target star has a planet and the
distribution of orbital periods and masses of planets.  Second, a
small number of observations are taken of each target star.  Bayesian
inference is used to calculate the posterior probability distribution
for all model parameters.  Then, the posterior probability
distribution for the model parameters is used to calculate the
predictive distribution, the posterior probability distribution for
the radial velocities at some future time.  By comparing the
information contained in the predictive distribution at various future
times, it is possible to choose observing times at which additional
observations would be most valuable.  This technique requires
performing integration over several variables and in general is
extremely computationally intensive.  In this paper we describe a
relatively fast algorithm for performing the necessary integrations.

%

While it would be desirable to observe each star in a
survey at the optimal times for each particular star, it is more
realistic to consider a planet survey which is able to perform a fixed
number of observations at specific times.  In this case, the
information contained in the predictive distributions for several
stars can be used to select which target star should be observed at a
given time.

We describe the algorithm for choosing observing times of a single
star in \S2.  In \S3 we describe a generalized algorithm that
allows for multiple target stars,
along with simulations which demonstrate the power of adaptive
scheduling algorithms.  In \S4 we present generalizations that allow
observing schedules to be optimized for meeting specific goals,
such as detecting planets in the habitable zone or maximizing the
number of planets detected.  Finally, we discuss the implications of
our findings and the challenges which remain in \S5.

\section{Adaptive Scheduling for a Single Target Star}
\label{SecSingleStar}

\subsection{Priors}
\label{SecPriors}

For each target star we assume a prior probability, $p(0)$, for the
null hypothesis that the radial velocity observations are consistent
with a constant radial velocity.  We assume a probability,
$p(1)=1-p(0)$, that the star has a single planet.  For the orbital
parameters of any planet, we take prior distributions that are flat in
$\log P$, $\log K$, and $\phi_o$, where $P$ is the orbital period, $K$
is the velocity semi-amplitude, and $\phi_o$ is the phase at a
given epoch.  These choices are standard for variables which represent
magnitudes and angles.  We limit the range of orbital parameters to
$P_{\min} \le P < P_{\max}$, $K_{\min} \le K < \infty$, and $0 \le \phi_o < 2
\pi$ (see \S2.3).  The choices for the prior distributions are supported by
scaling arguments as well as their approximate agreement with the
orbital parameters for the known extrasolar planets.

There are a few differences in our prior distributions and the
distributions of orbital elements of known extrasolar planets.  First,
we apply a sharp cutoff for orbital periods less than $P_{\min}$.  The
OGLE transit searches have discovered planets with orbital periods as
short as $1.2$d (Koanacki 2003).  While radial velocity searches are
very sensitive to such planets, the shortest orbital period discovered
by a radial velocity survey is 2.5d (Udry et al.\ 2003).  It is
important to recognize the OGLE transit search surveys a much larger
number of stars than radial velocity surveys (Gaudi, Seager, \&
Mallen-Ornelas 2004).  Thus, the observations imply that the
distribution of orbital periods is roughly flat in $\log P$ for $P \ge
3d$, but there is a very significant reduction in the number of
planets at shorter orbital periods (Gaudi, Seager, \&
Mallen-Ornelas 2004).  Therefore, it would be reasonable
to apply cutoff for orbital periods $P < P_{\min}$ for any $P_{\min} <
3$d, and we choose $P_{\min} = 2.5$d.  Given recent discoveries, we
do not suggest such a large $P_{\min}$ for future studies.


We also apply a sharp cutoff for orbital periods greater than
$P_{\max}$.  When a planet has an orbital period much longer than
$T_{\mathrm obs}$, then there are degeneracies in the Keplerian
orbital parameters and the radial velocities can be well modeled by a
quadratic polynomial (Cumming 2004).  By replacing all the Keplerian
models with $P \ge \pi T_{\mathrm obs}$ with a single quadratic model,
it is possible for our algorithm to detect planets with orbital
periods, $\pi T_{\mathrm obs} \le P$ efficiently.  We maintain the
flat prior in $\log P$ by setting the prior probability for the
quadratic model equal to the sum of the prior probabilities for
orbital periods in the range $\pi T_{\mathrm obs} < P < P_{\max}$.
Since the prior distribution is flat in $\log P$, this relative prior
probability is not sensitive to the exact choice of $P_{\max}$.  The
effect of varying $P_{\max}$ is to change the prior probability for a
planet having a long-period orbit relative to the prior probability
for a planet having an orbital period between $P_{\min}$ to
$P_{\max}$.  Since little is known about the abundance of extrasolar
planets with orbital periods greater than $\sim 10$yr, we choose
$P_{\max} = 40$yr guided by our own solar system.

Another difference between our prior distributions and the
distributions of orbital elements for known extrasolar planets is that
we assume the planetary orbits are circular, i.e., the orbital
eccentricity, $e$, is zero.  
Since a circular orbit approximates a Keplerian orbit with a small
eccentricity, our algorithm is expected to identify planets with small
and even moderately eccentric orbits.  However, the efficiency for
detecting planets on moderately eccentric orbits may be somewhat
reduced compared to the efficiency for detecting a planet on a
circular orbit with comparable mass and orbital period.  The
reduction in efficiency is relatively mild for $e<0.4$, but rapidly
becomes more significant (Endl. et al.\ 2002; Cumming 2004).
While many of the known extrasolar planets are on significantly
eccentric orbits, the planets with shorter orbital periods tend to
have smaller eccentricities.  This is likely due in part to tidal
circularization affecting planets with small orbital periods (Rasio,
Livio, \& Tout 1996).  While the assumption of circular orbits is
likely appropriate for many planets (especially short-period planets
targeted by the N2K project), there is no question that it would be
more desirable to include eccentricities.  Still, the assumption of
circular orbits permits significant computational advantages making it
extremely attractive when a large range of parameter space must be
searched (e.g., for planetary orbits which are poorly constrained by
the currently available data).  The reduction in computational
requirements makes it computationally feasible to explore the
properties of adaptive scheduling algorithms (as in this paper).

 \subsection{Initial Observations}

Before making observations of a target star, there is no basis for
believing that the star is more or less likely to have a planet or
that various orbital parameters are preferred, except what is
suggested by the prior probability distribution.  Therefore, an
initial set of $N_{\min}$ observations is made for each star.  When
the number of observations, $N_{\mathrm obs}$ is less than $N_{\min}$,
the choices for when to observe are not affected by previous
observations, however these choices may be affected by practical
considerations.  For example, observations must be made at night and
radial velocity surveys are typically allocated observing time near full
Moon.  Additionally, the airmass and atmospheric conditions in the
direction of each target may favor observing certain targets at
certain times during the available observing nights.  


Under the null hypothesis, there is a single fit parameter for the
constant velocity of the star, $C_0$, and the star's velocity is given
by $v_{*,C_0}(t) = C_0$.  Next, we consider the alternative hypothesis
that the star has a single planet in a circular orbit.  There are four
parameters which can be varied to fit the velocity observations of
each star, $P$, $K$, $\phi_o$, and $C_1$, where $C_1$ is the constant
velocity of the star.  (Given the way the star's velocity is measured,
it is typically necessary to use different values of $C_1$ for
different observatories.  Thus, when there are only a small number of
observations of a given target star, it is extremely advantageous if
all the observations are made from a single observatory.  For the
purposes of this paper, we assume that all radial velocity
observations are made with a single observatory.)  The radial velocity
signature of a planet on a circular orbit can be written as
\begin{equation}
v_{*,\vec{x}}(t) = K \cos \left[ \frac{2\pi}{P}t + \phi_o \right]  + C_{1}.
\end{equation}
%
%
After a set of $N_{\min}=3$ initial observations, it is possible to
evaluate the plausibility of the fit parameters, $P$, $K$,
$\phi_o$, and $C_1$.  Since each observation has some
observational uncertainties, the orbit still is not uniquely
determined, even if our general model is exactly correct.

As noted in \S\ref{SecPriors}, when a planet has an orbital period
much longer than the time span of observations, then the fit
parameters used above are not well determined by the radial velocity
observations.  Therefore, for orbital periods $\pi T_{\mathrm obs} \le P \le P_{\max}$, we model the radial velocity of the star as
\begin{equation}
v_{*,\vec{a}}(t) = a_0 + a_1 t + a_2 t^2,
\end{equation}
where $\vec{a} = \left( a_0, a_1, a_2 \right)$ is the set of
coefficients for the polynomial model.

\subsection{Inference}

\label{SecInference}

Once $N_{\mathrm obs} \ge N_{\min}$, we analyze the available
observations using the methods of Bayesian statistics after making
each new observation.  The results of the analysis can be used to make
informed choices for when stars should be targeted for additional
observations.  Let $\vec{d}$ denote the set of available data, in this
case the previous radial velocity observations.  We have already
introduced the prior probabilities for the null hypothesis, $p(0)$,
and for the single planet model, $p(1)$, as well as the prior
probability distribution for orbital parameters, $p(\vec{x}|1)$.
Next, we introduce the conditional probability for the observations
given the null hypothesis, $p( \vec{d} | 0 )$, and the conditional
probability for the observations given a fixed set of model
parameters, $p(\vec{d} | \vec{x} )$.  Since the observational errors
are assumed to be independent, both conditional probabilities can be
simply evaluated as the product of the probabilities for drawing each
observation given the relevant model for the stellar velocity.  Since
each radial velocity measurement is obtained by averaging the Doppler
shift measured for hundreds of spectra lines, the observational
uncertainties are very well approximated by a normal distribution, and
the conditional probabilities are given by
\begin{eqnarray}
p(\vec{d} | \vec{m} ) 
& = & \prod_i p(d_i | \vec{m}) 
= \prod_i \frac{\exp{\left[-\frac{\left(d_i - v_{*,\vec{m}}(t_i)\right)^2}{2\sigma_i^2} \right]}}{\sqrt{2 \pi} \sigma_i} \\\
& = & \frac{\exp \left[\frac{-1}{2} \sum_i \left(\frac{d_i-v_{*,\vec{m}}(t_i)}{\sigma_i}\right)^2 \right]}{\left(2\pi\right)^{N_{\mathrm obs}/2} \prod_i \sigma_i}
\equiv \frac{\exp \left[ \frac{-\chi^2(\vec{m})}{2}\right]}{\left(2\pi\right)^{N_{\mathrm obs}/2} \prod_i \sigma_i},  
\end{eqnarray}
where $\vec{m}$ represents the generalized model parameters, i.e.,
either $\vec{m}=(C)$ (the null hypothesis model), $\vec{m}=\vec{x}$ (the single
planet model with $P < \pi T_{\mathrm obs}$), or $\vec{m}=\vec{a}$ (the
polynomial model for a planet with $\pi T_{\mathrm obs} \le P \le
P_{\max}$).  Each individual observation, $d_i$, is made at a time,
$t_i$, and has an observational uncertainty, $\sigma_i$.  Since the
observational uncertainties are nearly Gaussian and assumed to be
independent, the conditional probability distribution for all the
available observations is a chi-squared distribution, and we
introduced the goodness of fit statistics, $\chi^2(\vec{m})$ which can be
easily computed for each set of model parameters, $\vec{m}$.

Next, we introduce terminology from Bayesian statistics, $p(\vec{d},
0)$, the joint probability for the observations and the null
hypothesis, and $p(\vec{d}, \vec{x})$, the joint probability for the
observations and the single planet hypothesis with a particular set of
model parameters, $\vec{x}$.  The joint probabilities can be written
as the product of the prior probability and the conditional
probability, e.g., $p(\vec{d}, 0) = p(0) p(\vec{d} | 0)$.  We will
also use Bayes theorem, which states that
\begin{equation}
p(\vec{m} | \vec{d}) 
 = \frac{p(\vec{d}, \vec{m})}{p(\vec{d})} 
 = \frac{p(\vec{m}) p(\vec{d} | \vec{m})}{\int d\vec{m} \, p(\vec{m}) p(\vec{d} | \vec{m})},
\end{equation}
We use the joint probabilities and Bayes' theorem, to compute the
posterior probabilities which incorporate both the prior probabilities
and the information contained in the observations, $\vec{d}$.  For
example, the posterior probability for the null hypothesis is
\begin{eqnarray}
p( 0 | \vec{d} ) & 
= & \frac{ p(\vec{d}, 0) }{p(\vec{d}, 0) + \int d\vec{x} \, p(\vec{d}, \vec{x}) + \int d\vec{a} \, p(\vec{d}, \vec{a}) } \\
& = & \frac{\int dC \, p(C) p(\vec{d}| C) }{\int dC \, p(C) p(\vec{d}| C)  + \int d\vec{x} \, p(\vec{x}) p(\vec{d} | \vec{x}) + \int d\vec{a} \, p(\vec{a}) p(\vec{d} | \vec{a}) }  \\
& = & \frac{p(0) \int dC \, p(C | 0) p(\vec{d}| C) }{p(0) \int dC \, p(C | 0) p(\vec{d}| C)  + p(1) \int d\vec{x} \, p(\vec{x} | 1 ) p(\vec{d} | \vec{x}) + p(1) \int d\vec{a} \, p(\vec{a} | 1 ) p(\vec{d} | \vec{a}) },
\end{eqnarray}
where $p(C | 0)$ is the prior distribution (flat) for the constant velocity
given the null hypothesis, $p(\vec{x} | 1)$ is the prior distribution
for the sinusoidal fit parameters given that a planet is present, and
$p(\vec{a} | 1)$ is the prior distribution for the polynomial fit
parameters given that a planet is present.

Unfortunately, the integrals, and particularly the integral over
$\vec{x}$ in the denominator, can be extremely difficult to evaluate.
In particular, $\chi^2(\vec{m})$ and hence $p(\vec{d} | \vec{x} )$
can be extremely ``bumpy'' functions (Ford 2005).  It is computationally
impractical to actually calculate $\chi^2(\vec{x})$ over the
entire range of parameter space with sufficient resolution to
approximate the integral accurately.  Therefore, we must find a way to
approximate the integrals in a computationally efficient manner.

When the orbital parameters are well constrained, then the integral is
typically dominated by the contribution from a small region of
parameter space near the best-fit solution and the integrals are
easily evaluated.  Even when the orbital parameters are somewhat less
constrained, the method of Markov chain Monte Carlo provides a
powerful tool for evaluating the necessary integrals.  However, even
Markov chain Monte Carlo is not computationally practical when the observations still
permit a wide variety of distinct orbital solutions (e.g., when there
are only a small number of observations).  Since the N2K
project is particularly interested in working with small data sets, we
expect that it will frequently be necessary to analyze radial velocity
observations which only provide limited constraints on the orbital
parameters, given the typical planetary masses, orbital periods, and
measurement errors.  Therefore we have developed an efficient
algorithm for approximating the necessary integrals.

%
While the integrand typically has numerous local maxima which can be
spread across a wide range of orbital periods, if the orbital period
is held fixed at $P$, then the integral over the remaining fit
parameters is dominated by the contribution from the single maximum
(assuming a circular orbit).  Therefore, we approximate the integral
by separating the integral over orbital period, $P$, from the
integrals over the remaining fit parameters, $\vec{x}_P$.  We sum the
contributions to the integral from each of the regions around the
best-fit solutions for each orbital period.  Thus, for the purposes of
computation, we replace the integral over orbital period with a
summation and will approximate the integrals over $\vec{x}_P$, giving
\be
p( 0 | \vec{d} ) = \frac{ p(0) \int dC \, p(C | 0) p(\vec{d}| C) }{p(0) \int dC \, p(C | 0) p(\vec{d}| C)  + p(1) \sum_i  \Delta \log P_i \int d\vec{x}_{P_i} \, p(\vec{x} | 1 ) p(\vec{d} | \vec{x}) + p(1) \int d\vec{a} \, p(\vec{a} | 1 ) p(\vec{d} | \vec{x})},
\ee
where $\Delta \log P_i$ is the spacing between the logarithm of successive orbital
periods, and $\vec{x}_P$ is the set of fit parameters excluding the
orbital period, $P$.  Similarly, the posterior probability for a planet with orbital period near $P$ is given by
\be
p( P | \vec{d} ) \Delta \log P = \frac{p(1) \Delta \log P \int d\vec{x}_{P} \, p(\vec{x} | 1 ) p(\vec{d} | \vec{x}) }{p(0) \int dC \, p(C | 0) p(\vec{d}| C)  + p(1) \sum_i  \Delta \log P_i \int d\vec{x}_{P_i} \, p(\vec{x} | 1 ) p(\vec{d} | \vec{x}) + p(1) \int d\vec{a} \, p(\vec{a} | 1 ) p(\vec{d} | \vec{x})},
\ee
and the posterior probability for a planet with an orbital period greater than 
$\pi T_{\mathrm obs}$ is
\be
p( P\ge \pi T_{\mathrm obs} | \vec{d} ) = \frac{p(1) \int d\vec{a} \, p(\vec{a} | 1 ) p(\vec{d} | \vec{x}) }{p(0) \int dC \, p(C | 0) p(\vec{d}| C)  + p(1) \sum_i  \Delta \log P_i \int d\vec{x}_{P_i} \, p(\vec{x} | 1 ) p(\vec{d} | \vec{x}) + p(1) \int d\vec{a} \, p(\vec{a} | 1 ) p(\vec{d} | \vec{x})  }  .
\ee
Clearly, the posterior probability that a star has a planet with any
orbital period is simply
\be
p( 1 | \vec{d} ) = \sum_i p( P_i | \vec{d} ) \Delta \log P_i + p(P\ge\pi T_{\mathrm obs}) = 1 - p( 0 | \vec{d} ).
\ee

For each orbital period, $P$, we must approximate each of the
integrals over $\vec{x}_P$.  Since the prior, $p(\vec{x} | 1, P_i)$ is
flat, we expand the argument of the exponential, $\chi^2(\vec{x}_P |
P)$, about its minimum ($P$ is held fixed).  Since we expand about a
minimum, the first derivatives of $\chi^2$ with respect to the
variable in $\vec{x}_P$ vanish.  Therefore the $\chi^2$ surface is a
quadratic function centered on the minimum, and we can approximate the
integral by extending the limits of integration to infinity.  The
resulting multidimensional Gaussian integral can then be evaluated
analytically, using only the value of $\chi^2$ at its minima,
$\min_{\vec{x}_P} \chi^2(\vec{x}_P | P)$, and the determinant of the
covariance matrix, $\mathrm{Covar}\left(\chi^2(\vec{x}_P | P)\right)$,
as
\be
\int d\vec{x}_{P} \, p(\vec{x}_P, P | 1 ) p(\vec{d} | \vec{x}) \simeq 
\frac{\sqrt{\mathrm{Det} \left|\mathrm{Covar}\left(\chi^2(\vec{x}_P | P)\right)\right|}}{\left(2\pi\right)^\nu \prod_i \sigma_i } \exp \left[ -\frac{1}{2} \min_{\vec{x}_P} \chi^2(\vec{x}_P | P)\right],
\ee
where $\nu = N_{obs} - N_{fit}$ and here $N_{fit}=3$  (Sivia 1996; Cumming 2004).  In a similar
way, the integral over $\vec{a}$ can be approximated by
\be
\int d\vec{a} \, p(\vec{a} | 1) p(\vec{d} | \vec{a}) \simeq 
\frac{\sqrt{\mathrm{Det} \left| \mathrm{Covar}\left(\chi^2(\vec{a})\right)\right|}}{\left(2\pi\right)^\nu \prod_i \sigma_i } \exp \left[-\frac{1}{2} \min_{\vec{a}} \chi^2(\vec{a})\right],
\ee
and the integral over $C$ can be approximated by
\be
\int dC \, p(C | 0) p(\vec{d}| C) \simeq
\frac{\sqrt{\mathrm{Var}\left(\chi^2(C)\right)}}{\left(2\pi\right)^\nu \prod_i \sigma_i } \exp \left[ -\frac{1}{2} \min_{C} \chi^2(C)\right],
\ee
where $N_{fit} = 1$ and the determinant of the covariance matrix has
been replaced by the square root of the variance of the single fit
parameter, $C$.  Thus, we approximate the necessary integrals by
explicitly summing the contributions from the null hypothesis, the
polynomial model, and all possible orbital periods, but approximate
the integrals over the remaining fit parameters as Gaussian integrals.
This provides a good approximation to the necessary integrals while
leaving only one dimension ($P$) which must be finely sampled.

It remains to identify the best fit solution for each orbital period
considered and to evaluate the probability of each of these possible
solutions.  This is equivalent to the problem of evaluating the
floating-mean periodogram.  The floating-mean periodogram and its
relationship to the standard periodogram is described by Cumming
(2004).  The periodogram is evaluated on a grid uniform in the
frequency, $f = 1/P$, rather than in $\log P$.  Thus, the factors
$\Delta \log P_i$ serve as weighting factors to ensure that we
maintain a prior which is uniform in $\log P$.  The necessary number
of orbital periods to consider is set by the ratio of the maximum
orbital period considered, $\pi T_{\mathrm obs}$, to the minimum
period considered, $P_{\min}$.  Since we do not want to miss a minima
in $\min_{\vec{x}_P} \chi^2(\vec{x}_P, P)$, we oversample by a factor
$\zeta \simeq 4$.  Thus, the number of orbital periods considered is
$N_P \simeq \zeta \pi T_{\mathrm obs} / P_{\min}$, a constant times
the Nyquist frequency corresponding to the minimum period.  Note that
we do not count sine and cosine components separately, and we are
searching for periods up to $\pi T_{\mathrm obs}$, despite the fact
that models with orbital periods longer than $T_{\mathrm obs}$ are so
similar that they can not be distinguished with the previous
observations.  The time span of observations can be as short as a
couple of months or extend for several years.  Hence it is typically
necessary to find the best-fit orbital parameters for thousands of
orbital periods.  Given the large number of global searches necessary,
the computation time required is significantly reduced if $\chi^2$ can
be written as a linear function of the fit parameters.  While this is
impossible for eccentric Keplerian orbits, it is possible for circular
orbits by writing
\begin{equation}
v_{*,\vec{x}}(t) = A \cos (\frac{2\pi}{P}t) + B \sin (\frac{2\pi}{P}t) + C_{1},
\end{equation}
where $K = \sqrt{A^2+B^2}$, and $\phi_o = \tan \frac{B}{A}$.  Using
this formulation allows for each of the best-fit solutions and the
covariance matrices to be evaluated by linear least-squares which is
much faster and more robust than non-linear least squares.  Since we
used $A$ and $B$ as fit parameters rather than $\log K$ and $\phi_o$,
we must include a weight equal to the determinant of the Jacobian of
transformation, $\left|J\right|=K^{-2}$.  While the Jacobian should
formally be inside the integral, we approximate the integral by
substituting the value of $K$ at the minimum in $\chi^2(\vec{x}_P |
P)$.  

While $K$ is allowed to take on any positive value, for the purpose of
comparing the posterior probability of the no-planet and one-planet
models it is necessary to normalize the prior distribution for $K$.
For this purpose only, we assume $K_{\min} \le K \le K_{\max}$, where
$K_{\max}$ is the amplitude of a 10 Jupiter-mass planet orbiting a solar-mass
star with an orbital period of $P_{\min}$.  Again, solely for the purposes
of setting the normalization, we adopt $K_{\min}$ is the signal amplitude for
which there would be a $\sim50\%$ probability of detecting the planet
and $K_{\max}$ is the maximum velocity amplitude of a planet for the
specified orbital period.  For a planet on a circular orbit, $K_{\max}
= 2\pi (m_{\max}/M_*).\left( G (M_*+m_{\max}) / P\right)^{-1/3}$,
where $m_{\max}/M_* = 0.01$ is the ratio of the maximum planet mass to
the star mass and $G$ is the gravitational constant.  Note that
$K_{\max}$ and hence the normalization for the prior, $p(\vec{x})$,
varies with the orbital period.  For $K_{\min}$ we use the analytic
approximation from (Cumming et al.\ 2002), $K_{\min} = 2
\sigma_{\mathrm obs} \sqrt{T_{\mathrm obs} / \left( P_{\min}
(N_{obs}-3) FAP \right)}$ , where $\sigma_{\mathrm obs}$ is the
uncertainty of the individual velocity measurements, $N_{obs}$ is the
number of previous observations of the star, and $FAP$ is the false
alarm probability which we set to $1/N_{\mathrm targets}$, the inverse
of the number of target stars.  As pointed out by an anonymous
referee, this choice is somewhat arbitrary.  Nevertheless, given our
choice of prior, it is necessary to make some choice to result in a
normalized probability distribution.  In the future, we would suggest
using a single normalized prior distribution that has support below
$K_{\min}$, such as the modified Jeffreys prior, $p(K) = (K+K_o)^{-1}
/ \log\left[1+K/K_o\right]$.

The above procedure allows us to efficiently calculate the probability
of the null hypothesis as well as a list of probabilities that there
is a planet with each of the orbital periods considered.  For each of
these probabilities, there is also a set of best-fit parameters and a
covariance matrix which describe the size and shape of the posterior
probability distribution for the remaining fit parameters.  To the
extent that our model and approximations are valid, these
probabilities and covariance matrices provide the optimal basis for
making inferences about the presence of a planet and its orbital
parameters.  Each time that a new observation is made, the entire procedure
is repeated to produce updated posterior distributions which
incorporate the new information from the latest observation.

\subsection{Prediction}
\label{SecPrediction}

Having estimated the posterior probability distributions for model
parameters in \S\ref{SecInference}, it is straightforward to sample
from $p(v(t) | \vec{d})$, the predictive probability distribution for a
hypothetical radial velocity observation at time, $t$.  
\be
p(v(t) | \vec{d} ) 
= \int d\vec{m} \, p(v(t) | \vec{m}) p(\vec{m} | \vec{d} )
= \int dC p(v(t)|C) \, p(C| \vec{d})
+ \int d\vec{x} \, p(v(t)|\vec{x}) p(\vec{x} | \vec{d})
+ \int d\vec{a} \, p(v(t) | 1, \vec{a}) p(\vec{a} | \vec{d})
\ee

Various summary statistics can be computed for $p(v(t) | \vec{d})$ at
each of several possible future observing times.  Perhaps the
simplest is to calculate the mean ($E[v(t) | \vec{d}]$) and variance
($E[\mathrm{Var}(v(t) | \vec{d})]$) of the velocities sampled from
$p(v(t) | \vec{d})$.  This can be done extremely efficiently and has
the added benefit that that the mean and variance are straight forward
to interpret.  

Naively, it might seem desirable to make future observations when
$E[\mathrm{Var}(v(t) | \vec{d} )]$ is largest (See Fig.\ 1).  While this seems to be a
reasonable strategy, a more rigorous analysis will lead to a somewhat
different result, as demonstrated in \S\ref{SecDesign}.

\subsection{Design}
\label{SecDesign}

A more sophisticated analysis incorporates the concept of a utility
function from decision theory.  
The utility function makes explicit
the utility of a specific combination of an action (e.g., observe at
time $t$) and an outcome (e.g., measure a velocity $v(t)$).  While the
experimenter can choose the action to be taken, the outcome is not
known a priori.  Nevertheless, the predictive distribution, $p(v(t) |
\vec{d})$, contains information about likelihood of various outcomes
for a given action.  Thus, the experimenter can calculate the expected
value of the utility function for various possible actions.  Then the
action with the largest expected utility can be chosen.  
Throughout this section, we closely follow the derivation of Loredo (2004).


While numerous utility functions are possible, one particularly well
motivated choice is to set the utility function equal to the change in
the information contained in the posterior probability distribution
for model parameters after incorporating the future observation.  Let
$I\left\{f(z)\right\}$ be the information contained in the distribution
$f(z)$, which is the negative of the Shannon entropy and is given by
\be
I\left\{f(z)\right\} = \int dz \, f(z) \log f(z).
\ee
The expectation for the information contained in the posterior
distribution for the model parameters after incorporating the future
observation is
\be
E[I\left\{p(\vec{m} | \vec{d}')\right\} ] = \int dv \, p(v|\vec{d}) I\left\{p(\vec{m} | \vec{d}')\right\},
\ee
where $\vec{d}'$ is the set of previous observations, $\vec{d}$,
augmented by the future observation, $v$.  Next, we will invoke Shannon's theorem which can be derived by considering the information contained in a joint probability distribution, in this case, $p(\vec{m},\vec{d}')=p(\vec{m}|\vec{d}')p(\vec{d}') = p(\vec{d}'|\vec{m})p(\vec{m})$.   By writing out the integrals contained in
\be
I\left\{p(\vec{m}|\vec{d}') p(\vec{d}') \right\} = I\left\{p(\vec{d}'|\vec{m}) p(\vec{m}) \right\},
\ee
separating integrals when possible, and simplifying integrals over probability densities which integrate to unity, we arrive at Shannon's theorem,
\be
I\left\{p(\vec{m} | \vec{d})\right\} + \int d\vec{m} \, p(\vec{m} | \vec{d} ) I\left\{ p(v|\vec{m}) \right\} = 
I\left\{p(v | \vec{d} )\right\} + \int dv \, p(v | \vec{d}) I\left\{p(\vec{m} | \vec{d}') \right\},
\ee
to rewrite the expected information as
\be
\label{EqnExpInfo}
E[ I\left\{p(\vec{m} | \vec{d}')\right\} ] =
I\left\{p(\vec{m} | \vec{d})\right\} + \int d\vec{m} \, p(\vec{m} | \vec{d} ) I\left\{ p(v|\vec{m}) \right\} 
- I\left\{p(v | \vec{d} )\right\}.
\ee
The first term in Eqn.\ref{EqnExpInfo} is simply the information about
the model parameters already available from the previous observations,
$\vec{d}$, and is independent of future observations.  The second term
is the weighted average of the information contained in the
probability distribution for the future observation conditioned on a
particular model.  Note that the distribution, $p(v(t)|\vec{m})$, is
the distribution of the observed velocities if the model where exactly
known.  Although the location of the distribution for the predicted
velocities depends on $t$ and $\vec{m}$, the shape and scale of the
distribution is independent of both $t$ and $\vec{m}$.  Since the
Shannon entropy of a distribution depends only on the scale and shape
of a distribution (and not the location where it is centered), the
second term of Eqn.\ \ref{EqnExpInfo} is also constant.  The remaining
term is the information content of the predictive distribution,
$p(v(t)|\vec{d})$, and has an explicit dependence on $\vec{d}$.  Thus,
the expected change in the information content of the posterior
probability distribution for the model parameters is
\be
E[ \Delta I\left\{p(\vec{m} | \vec{d}')\right\} ](t) = I\left\{ p(v(t) | v_{*}(t)) \right\} 
- I\left\{p(v(t) | \vec{d} )\right\},
\ee
where $v_{*}(t)$ is the actual radial velocity of the star at time $t$ (as opposed to the observed velocity $v(t)$).  The first term depends only on the distribution for the measurement about the true value, which we assume is independent of time.  Therefore, the expected
change in the information is maximized if the next observation is
taken when the information content of the predictive distribution is
minimized and the entropy of the predictive distribution is maximized.
Thus, an observing program will more efficiently constrain the orbital
parameters of a given target star if future observations are made at
times when the uncertainty in the predictive distribution is large.

The above analysis naturally leads to the technique of maximum entropy sampling.
Once $p(v(t) | \vec{d})$ has been estimated as outlined in sections
\ref{SecInference}, the Shannon entropy, $-I\left\{p(v(t) | \vec{d})\right\}$ can be
easily calculated for numerous possible future observation times.  
In particular, the necessary integral can be written as
\be
I\left\{p(v(t)|\vec{d}) \right\} = \int d\vec{m} \, p(\vec{m}|\vec{d}) \int dv p(v|\vec{m}) \log \left[ \int d\vec{m}' p(\vec{m}'|\vec{d}) p(v|\vec{m}') \right],
\ee
where the first integral represents a sum sampling the model
parameters from $p(\vec{m}|\vec{d})$, and the second integral
represents a sum sampling the prospective velocity from $p(v|\vec{m})$
using the previously drawn model parameters.  For each velocity drawn
in this manner, we must calculate the probability of obtaining that
velocity according to the full posterior distribution as the argument
to the logarithm.

By choosing to make the next observation when $I\left\{p(v(t) |
\vec{d})\right\}$ is near a minimum, the observation is expected to
yield more information about the model parameters than if the next
observation time were chosen randomly.  Once a new observation is
made, we must repeat the entire process of calculating a posterior
probability distribution for the model parameters, the predictive
distribution for the velocity at future times, and the entropy of the
predictive distribution at each time.

\subsection{Maximum Entropy versus Maximum Variance}

It is easy to demonstrate that the Shannon entropy of a Gaussian
distribution with standard deviation, $\sigma$, is $-\log (\sqrt{2\pi
e} \sigma)$.  If the uncertainty in the prospective measurement is
Gaussian with variance $\sigma_{\mathrm obs}^2$ and the predictive
distribution is also well approximated by a normal distribution with
variance $\sigma_{\mathrm pred}^2(t)$, then the expected change in information
reduces to
\be
E\left[ \Delta I\left\{p(\vec{m} | \vec{d}')\right\}(t)\right] = \log\left(\frac{\sigma_{\mathrm pred}(t)}{\sigma_{\mathrm obs}}\right).
\ee
Thus, the more simplistic strategy of choosing observation times to
maximize the variance (rather than the entropy) of the predictive
distribution (as described in \S\ref{SecDesign}) is equivalent when
the predictive distribution is normal.  Based on visual inspection of
several predictive distributions, we observe that the predictive distribution 
is typically well approximated by a normal distribution, if the period is assumed
to be known precisely.  While the predictive
distribution may be well approximated by a Gaussian distribution for
well constrained orbits, when $N_{\mathrm obs}$ is small, the
predictive distributions are generally not well approximated by a
Gaussian.  In particular, if there is a significant probability for
two qualitatively different models (e.g., null hypothesis and a planet
with orbital period near $P$), then the predictive distribution is
frequently bimodal with one mode centered on the best-fit constant
velocity and another mode near the velocity predicted by the best-fit
sinusoidal solution with a different orbital period.  
For example, let us
consider a case where there is a posterior probability, $p_a$, for 
models with orbital period near $P_a$ and predictive velocity
distribution approximately Gaussian centered on $v_a(t)$ with standard
deviation $\sigma_a(t)$, and there is a posterior probability, $p_b$,
for models with orbital period near $P_b$ (or perhaps the null model)
and a predictive distribution approximately Gaussian centered on
$v_b(t)$ with standard deviation $\sigma_b(t)$.  The the predictive
distribution is approximated by
\be
p(v(t)|\vec{d}) \simeq p_a N(v_a(t), \sigma_a(t)) + p_b N(v_b(t), \sigma_b(t)),
\ee
and the information contained in the predictive distribution is approximated by
\be
I\left\{p(v(t)|\vec{d})\right\} \simeq 
p_a I\left\{N(v_a(t),\sigma_a(t))\right\} + p_b I\left\{N(v_b(t),\sigma_b(t))\right\} \simeq 
0.5 p_a \log\left(2\pi e \sigma_a\right) + 0.5 p_b \log\left(2\pi e \sigma_b\right),
\ee
if $v_a(t)-v_b(t) \gg \sqrt{\sigma^2_a(t) + \sigma^2_b(t)}$.  Most
notably, the information in the predictive distribution is not
sensitive to the separation $v_a(t)-v_b(t)$, but the variance in the
distribution obviously does depend on the separation $v_a(t)-v_b(t)$.

As can be seen in this example, the entropy of such a distribution is not
sensitive to the separation between the modes, unlike the variance
which increases with the separation between the two modes.  Thus,
choosing observing times based on the variance rather than the entropy
of the predictive distribution will tend to favor observing at times
when the observations do not completely rule out another possible
model which predicts a very different velocity.  While choosing future
observation times based on the variance of the distribution may be
acceptable for well constrained orbits, it is particularly important
to use maximum entropy sampling when the observations are not yet able
to exclude qualitatively different models.

\subsection{Examples}
\label{SecExamples}

We have begun to apply the inference and predictive steps to some of
the observations taken near the beginning of the N2K project.  In
Fig.\ 1, we show the expected velocity and the 5th and 95th
percentiles of the predictive distribution as a function of time for
several target stars.  We show the median of the predictive
distribution as the heavy line and the credible intervals as the
thinner lines.  

By inspecting confidence intervals for the predictive distributions based on
actual observational data, we have identified four common cases.  
\begin{enumerate}
\item There is structure in the predictive distribution both during
the prior observations and significantly after the last observation.
The orbital period is well constrained and the structure is nearly
periodic with the same period (e.g., Fig.\ 1, top row).  

\item There is structure in the predictive distribution both during
the prior observations and significantly after the last observation.
The orbital period is not precisely known, and so the structure is not
periodic or is nearly periodic on a timescale significantly longer
than the orbital period (e.g., Fig.\ 1, lower left).  

\item In other cases, there is significant variability in the scale of
the predictive distribution times in the past, but not for
times in the future (Fig.\ 1, lower right).  This can occur when the
orbital period is only weakly constrained.  The uncertainty in the
orbital period causes information about the orbital phase to be lost
with time and the uncertainty in the orbital phase dominates
the width of the predictive distribution.  

\item There is no significant variability in the predictive
distribution around the prior observations, but the variability grows
with time due to the possibility of a long period planet (polynomial
terms).
\end{enumerate}

In the first two cases, maximum entropy sampling could provide a
valuable increase in the efficiency of constraining orbital
parameters.  In the first case, the structure in the predictive
distribution is periodic, so it is possible to identify the best time
to observe the system each orbital period.  However, in the second
case, it is not clear how frequently the system should be observed.
If the system is observed each time there is a local maximum in the
variance or entropy of the predictive distribution, then many
observations may be made during a single orbital period.
Alternatively, if the system is not observed at each local maximum,
then observations may skip an entire orbital period.  The last two
cases illustrates another problem with the maximum entropy sampling
algorithm applied to a single star.  When the entropy of the
predictive distribution increases with time, then maximum entropy
sampling does not identify any particular time.  In the next section
we present a variation on this algorithm which overcomes these
difficulties.

\section{Adaptive Scheduling for Multiple Target Stars}
\label{SecMulti}

In modern planet searches, there is typically a large list of possible
target stars and another list of opportunities to observe a small
subset of these stars.  Here we rephrase the goal of adaptive
scheduling algorithms.  Instead of identifying the best time to
observe a particular target, we ask which targets would be best to
observe at a particular time.  In this context, an adaptive scheduling
algorithm determines both the times at which each target star is
observed and the number of times each target star is observed.
Adaptively choosing the observing times as in \S\ref{SecSingleStar}
can significantly improve the efficiency for constraining orbital
parameters for stars with planets as demonstrated by Loredo (2004).
Similarly, adaptively choosing the number of observations of each
target star can significantly improve the efficiency for detecting
planets.  Thus,adaptive scheduling algorithms can provide a double
benefit to planet searches.

\subsection{Maximum Entropy}
\label{SecMultiMaxEntropy}

A straightforward generalization of the methods described in
\S\ref{SecSingleStar} is to apply the principles of maximum entropy
sampling to the joint posterior distribution function for model
parameters for each target star, $P \equiv p(\vec{m}_1, \vec{m}_2, ...,
\vec{m}_{N_{\mathrm targets}} | \vec{d}_1, \vec{d}_2, ...,
\vec{d}_{N_{\mathrm targets}})$, where $\vec{m}_i$ are the model
parameters for the $i$th star and $\vec{d}_i$ are the observations of
the $i$th star.  Since the posterior distributions for the fit
parameters of each star are independent of each other, $P = \prod_i p(\vec{m}_i | \vec{d}_i)$.
Therefore, the information contained in the joint posterior
distribution is simply the sum of the information contained in the
posterior distribution of each star independently.  
\be
I\left\{ P \right\} = \sum_i I\left\{ p(\vec{m}_i | \vec{d}_i ) \right\},
\ee
and the expected
increase in information about the joint distribution is equal to the
expected increase in information about the posterior distribution of
the star being targeted, 
\be
E\left[ \Delta I\left\{P\right\} \right](t,i) = E\left[\Delta I\left\{p(\vec{m}_i | \vec{d}_i)\right\}(t)\right],
\ee
where $i$ indicates which star is being targeted at time $t$.  Thus,
one can calculate the expected increase in information about the
posterior distribution for the model parameters for each star
separately and then choose to observe the star which is expected to
yield the most information.  

After each new observation is obtained, the above procedure can be
repeated and a new target star chosen.  In
practice, the procedure that we describe requires significant
computation time and dozens of radial velocity observations are made
on a clear night.  Since the orbital periods (and hence timescale for
variability in the predictive distributions) are typically long
compared to one night of observing, it is reasonable to calculate the
predictive distributions and entropy for each star at one time during
a night of observing (e.g. the time at which the star reaches its the
maximum altitude during the night).  A list of the stars with the largest
expected increase in information can be targeted during that observing
night. 

In principle, the predictive distributions and entropy for each star
could be calculated at several times during a night of observing.
This would make it possible to choose the observing time precisely,
rather than just which night to observe the star.  This could be
valuable for stars with very short orbital periods or highly eccentric
orbits (and hence short timescales for periastron passage).  In
practice, there are significant limitations on when a star can be
observed and costs associated with observing stars in an arbitrary
order.  (We will discuss how to incorporate this costs in
\S\ref{SecAltCosts}.)  For simplicity, in our simulations described below we do not
attempt to optimize the observing schedule across times within one
night.

\subsection{Example}
\label{SecMultiExample}

To demonstrate the value of adaptive scheduling, we have simulated
radial velocity planet surveys using both regular and adaptive target
scheduling.  First, we generate a list of 1000 target stars and
randomly assign planets to some of them.  The frequency, mass, and
orbital period distributions are taken from Tabachnik \& Tremaine
(2002).  We then randomly choose 20 observing nights per year.  To simulate
the allocation of nights on a large telescope, we restrict the
possible observing nights to be during the quarter of the lunar month
closest to full moon.  Each night 100 observation times are regularly
spaced during the night.  For the regular scheduling algorithm, stars
with the smallest number of observations are given the highest
priority.  Among stars which have the same number of observations, the
stars which are less frequently observable are given priority.  For
the adaptive scheduling algorithm, each star is observed three times
as with the regular scheduling algorithm.  Subsequently, a Bayesian
analysis (as described in \S\ref{SecInference}) of the available
observations is performed at the conclusion of each night a star is
observed.  Before each observing night the predictive probability
distribution is calculated (as described in \S\ref{SecPrediction}) for
the velocity of each star observable on that night.  The exact time
for calculating the predictive distribution is the time at which the
star reaches maximum altitude during the night.  The possible target stars are
prioritized based on the entropy of the predictive distribution.  The
100 stars with the highest priorities are observed that night in order of their
right ascension (not necessarily at the exact time for which the
predictive distribution was calculated).

Here we present a summary of the results of these simulations.  At the
end of each night we monitor the posterior distributions for the model
parameters of each system, paying particular attention to the number
of planet detections (which we define to be systems for which the
probability of the null hypothesis, no planet, is less than 0.1\%).
%
%
In Fig.~2, we show how the number of detections
increases as a function of the number of observing nights.  The
adaptive scheduling algorithm based on \S\ref{SecMulti} (dashed blue)
%
%
is clearly more efficient than the regular scheduling algorithm (solid
black) for detecting planets, even though it is not explicitly
optimized for detecting the largest number of planets.  More
importantly, the additional planets that are being detected by the
adaptive scheduling algorithm, tend to be those with the smallest
velocity amplitudes (see Fig.~3).  This is accomplished by observing some stars
more frequently than others.  In Fig.~4 we present a histogram showing
the fraction of stars that were observed a given number of times.
While the regular scheduling algorithm observed each star ten times,
the adaptive algorithms observed many stars slightly less frequently
and a few stars much more frequently.  This makes the adaptive
scheduling algorithms much more sensitive to planets with velocity
amplitudes near the threshold of detection.  Thus, while the total
number of planets detected increases by $\sim10-20\%$, the mass of
the least massive planet detected by the adaptive scheduling algorithm
is less than that of the regular scheduling algorithm by a factor
$\sim2$ or more.  It is also important to note that the accuracy with
which the orbital parameters are measured has not been sacrificed (see Fig.~5).
While the orbital periods and amplitudes for planets with the largest
velocity amplitudes ($\ge100$m/s) are measured with a similar
accuracies, the adaptive scheduling algorithm provides a significant
improvement in the accuracy of the orbital parameter determinations
for planets with more modest velocity amplitudes ($\le30$m/s).

\section{Alternative Utility Functions}
\label{SecAltUtility}

In \S\ref{SecSingleStar} we focused on when to observe a single star, and
hence the predictive distribution, $p(v | \vec{d})$, and its entropy
were an obvious choices for comparing the utility of observations at
various times.  In \S\ref{SecMulti}, we focused on choosing which
star (from a large list) should be targeted at the next observing
opportunity.  In \S\ref{SecMulti}, we choose a utility function
based on the joint posterior for the model parameters for all the
target stars.  However, in this case the choice of utility function is
less obvious.  
Various surveys and investigators may have differing goals and hence
differing utility functions.  For example, one possible goal would be
to measure the orbital parameters to some desired accuracy.  Another
reasonable goal might be to discover as many planets as possible given
some fixed amount of observing time.  In that case, it would make
sense to stop observing stars once it had been established that they
harbored a planet, even if the orbital elements were not yet well
constrained.   An even more extreme example is for a radial velocity
survey intended to help select reference stars for astrometric survey
by future missions such as SIM.  In that case, one could stop
observing a star before obtaining a rough measure of the orbital
parameters or even before the false alarm rate (for detecting a
planet) was small.  This is similar to a strategy for a survey aimed at
discovering planets with a small mass which would eliminate stars once the
radial velocities are observed to vary over too large a range to be
due to a low mass planet.  Yet another possible goal would be to
discover multiple planet systems or planets with long orbital periods.
In this case, one would not want to stop observing a star even after
the orbit of one planet had been well characterized, if it was still
possible the system could have an additional planet with a longer
orbital period.

The above examples illustrate that simply targeting stars based on the
maximum entropy method with the same utility function is not always
the best strategy for a given application.  Nevertheless, we have
demonstrated that adaptive scheduling algorithms can significantly
increase the efficiency of an observing program.  Thus, it is
important that the goals of an observing program be carefully
considered and clearly identified.  Then, a utility function can be
chosen that is appropriate for the particular purpose of the
observations.  Once a utility function has been defined, the methods
outlined in this paper can be used to optimize the observing program
for the given utility function.  

The utility function discussed above, $E\left[\Delta
I\left\{p(\vec{m}|\vec{d}')\right\}\right]$, is relatively easy to
calculate based on the posterior distribution, $p(\vec{m}|\vec{d})$,
giving it a practical advantage over many other possible utility
functions.  In this section we describe simple generalizations of the
above utility function which are can be computed with a similar
efficiency.  The generalized maximum entropy utility functions that we
describe below provide a means for optimizing observing schedules for
a broad range of goals.
%

\subsection{Information about a Subset of Models}
\label{SecMultiAltSubset}

One case worth considering is when we are only interested in models
which satisfy certain criteria.  For example, we might only be
interested in obtaining more information about stars with planets (and
not about the constant velocity of stars without planets).  Similarly,
we might be interested in companions with (minimum) masses less than
some threshold, perhaps to exclude binary stars or perhaps to target
terrestrial-mass planets.  In this case we could replace
$I\left\{p(\vec{m})\right\}$ with
\be
I_{\Theta}\left\{p(\vec{m})\right\} = \int d\vec{m} \, p(\vec{m}) \theta(\vec{m}) \log p(\vec{m}),
\ee
where $\Theta$ is a region of the model parameter space,
$\theta(\vec{m}) = 1$ when the parameters $\vec{m}$ satisfy a certain
criteria and $\theta(\vec{m}) = 0$ otherwise.  The expression for
$I_{\Theta}(p(\vec{m}))$ can be thought of as the information contained in
the distribution $p(\vec{m})$ about the subset of model parameters in $\Theta$.  In this case, the relevant utility function becomes
\be
E\left[\Delta I_\Theta\left\{p(\vec{m}|\vec{d}')\right\}\right] = p(\vec{m} \in \Theta) I\left\{p(v|v_{actual})\right\} - \int d\vec{m} \, p(\vec{m}|\vec{d}) \theta(\vec{m}) \int dv \, p(v|\vec{m}) \log \left[ p(v|\vec{d}) \right],
\ee
where the first term is again a constant, provided that the scale and
shape of the sampling distribution does not depend on the model
parameters.

In the above example, we used $\theta(\vec{m})$ as an indicator
variable to specify when the parameters satisfied some criteria of
interest, such as whether the model includes a planet or whether the
orbital period is less than some threshold.  More specialized forms of
$\theta(\vec{m})$ could be chosen for specific goals, such as finding
planets with certain orbital periods (e.g., within the habitable
zone).  In principle, $\theta(\vec{m})$ could be used as a weight,
specifying the relative value of information about systems with
various model parameters.  For example, a planet search aiming to
discover low-mass planets could specify a $\theta(\vec{m})$ which
decreases for high mass planets.

\subsection{Information in Marginal Distributions}
\label{SecMultiAltMargin}

Another case worth considering is when some model parameters are of
more scientific interest than others.  For example, the constant
stellar velocities, $C_0$ and $C_1$, contain no information about
extrasolar planets.  Similarly, the angle $\phi_o$ may be of less
interest than other model parameters such as the orbital period, $P$.
As an extreme example, one might be interested in determining only if
a star has a planet and not be interested in measuring the orbital
parameters.  In such cases, it is useful to subdivide the set of model
parameters, $\vec{m}$, renaming them $(\vec{m},\vec{n})$, where
$\vec{m}$ is the set of scientifically interesting model parameters
and $\vec{n}$ is the set of ``nuisance'' parameters.  Now, we can
marginalize over the nuisance parameters and consider the expected
change in the information contained in the probability distribution,
$p(\vec{m}|\vec{d}') = \int dn p(\vec{m}, \vec{n} | \vec{d}')$, rather
than using the joint distribution $p(\vec{m}, \vec{n} | \vec{d}')$
as before.  Thus, the relevant utility function becomes
\be
E\left[\Delta I\left\{p(\vec{m}|\vec{d}') \right\} \right] = \int d\vec{m} \, p(\vec{m}|\vec{d}) \int d\vec{n} \, p(\vec{n}|\vec{d},\vec{m}) \int dv \, p(v|\vec{m},\vec{n}) \log \left[ \frac{ \int d\vec{n}' p(\vec{n}'|\vec{d},\vec{m}) p(v|\vec{m},\vec{n}') }{\int d\vec{m}' \int d\vec{n}' p(\vec{m}',\vec{n}'|\vec{d}) p(v|\vec{m}',\vec{n}')} \right].
\ee
Here the first two integrals sample over all possible models and the
third integral samples over all possible values of the observed
velocity, just as before (e.g., Eqn.\ 24).  Indeed, if the logarithm
were split into the difference of two terms, then the term arising
from the denominator would also be mathematically equivalent to the
(negative of) Eqn.\ 24.  However, the term arising from numerator
causes this utility function to differ from Eqn.\ 23, as it no longer
simplifies to equal the information of the upcoming observation if the
actual velocity were known.  For the utility function in Eqn.\ 23, the
entropy of the predictive distribution at the time of a hypothetical
future observation is compared to the entropy of the probability
distribution for the observed value given the actual value.  However,
for this choice of utility function, the entropy of the predictive
distribution at the time of a hypothetical future observation is
compared to the entropy of the predictive distribution marginalized
over the nuisance parameters (evaluated for the same time).  Thus,
both terms depend on the time, and a hypothetical future precise
observation would be expected to contribute less information at times
where the predictive distribution is more sensitive to the values 
of the nuisance parameters.  

If we marginalize over all the fit parameters, then we can obtain
posterior distributions for the probability that the system does or
does not have a planet.  A scheduling algorithm based on maximizing
the expected increase in information contained in this distribution is
expected to detect planets very efficiently.  Indeed, we can see that
this is the case from the dotted red curve in Figs.~2 \& 3.

\subsection{Non-Greedy Algorithms}
\label{SecMultiAltNonGreedy}

So far we have restricted our attention to ``greedy'' algorithms,
since the utility functions have only considered the effect of a
single additional observation (Cormen et al.\ 2001).  As we have demonstrated, these
greedy algorithms perform quite well in the cases which we considered.
However, it is worth briefly discussing alternative ``non-greedy''
utility functions.

Let us consider the case where making one additional observation is
expected provide no or little increase in information, but making
multiple additional observations (perhaps at particular times) would
be expected to provide significant additional information.  A simple
example of such a situation is when a new target star is added to the
survey.  The first two observations of any star can not result in the
detection of a planet.  Yet, if the current target list has been
searched thoroughly, then it could be more productive to add new
stars to the target list.  While this effect is increased when the
constant velocities are considered nuisance parameters, there is
some effect even when $C_0$ and $C_1$ are considered parameters of interest.

In principle, cases such as these can be handled by calculating the
expected increase in information at some later time by which several
additional observations could be made.  Multiple additional
observations are considered by sampling over the various possible
combinations of future observations.  This generalization has the
obvious drawback that it is necessary to perform an additional
integral over various possible combinations of observing schedules.
One possible simplification is to relax the assumption that the total
number of observations is held fixed (only for the purpose of
evaluating the integral over future observing schedules) and to
consider observing schedules for each star separately.  Then the integral
over future observing schedules can be performed separately for each
star by assuming a constant probability of making an observation at
each possible future observing time.

In principle, the above algorithm could identify combinations of
multiple observations which are significantly more valuable than would
be estimated by a greedy algorithm.  Another benefit of the above
algorithm is that it can automatically account for differences in the
fraction of time during which stars are observable (e.g., due to
seasonal effects).  We have performed a few small simulations using
the non-greedy algorithm described above and found that they provided
a small increase in observing efficiency for the cases which we
considered.  However, a through comparison of greedy and non-greedy
algorithms will require significantly more computation power.

\subsection{Utility Functions with Costs}
\label{SecAltCosts}

The utility function can also include information about the cost of
making a given observation.  While economic cost is often used in
Bayesian decision theory, for our applications it is preferable to use
observing time necessary to perform an observation as the measure of
cost.  The observing time required for a given observation includes
the time necessary to collect the desired number of photons, but also
the time required for CCD readout and slewing the telescope from the
previous target to the next target.  The time required for CCD readout
is known and a constant for each exposure.  (For faint target stars,
it may be necessary to use multiple exposures for a single
observation, if the barycentric correction would vary significantly
during the necessary integration time.)  For target stars near the
previous target, the telescope can often slew to the target during CCD
readout, resulting in no or minimal loss of observing time to slewing.
Typically, observing schedules within a night are chosen so that the
vast majority of the observing time is spent integrating on targets.
The main factor in determining the integration time necessary is the
apparent magnitude of the target star.  In principle, adaptive scheduling algorithm could plan an entire evenings
observations, considering the cost to observe each possible target
throughout the night.  However, the exact amount of
integration time depends on the atmospheric extinction (which we
assume in proportional to the airmass) as well as time variable atmospheric
conditions that are generally not known in advance.
Therefore, we do not believe that it is worthwhile to include such a
fine level of scheduling when selecting targets for a given night.  In
practice, atmospheric conditions (e.g., seeing and cloud cover) change
throughout a night, making it impossible to know in advance the number
of targets that can be observed during that night.  Given the
computational complexity of these algorithms, it is impractical to
perform new analyzes throughout the night as atmospheric conditions
change.  By assuming that atmospheric extinction is a function of
airmass only (e.g., ignoring the possibilities of cloud cover or
atmospheric conditions changing throughout the night), we can compute
the expected amount of observing time necessary for each possible
target star.  Then, we identify the targets with the greatest expected
increase in information per unit observing time (rather than per
observation).


\section{Discussion}
\label{SecDiscussion}


We have developed a practical algorithm for applying adaptive
scheduling to radial velocity planet searches.  The algorithms
presented are rigorously grounded in Bayesian data analysis and
information theory, and still permit specialization for the specific
goals of the observing program.  While such adaptive scheduling
algorithms are computationally demanding, they can provide dramatic
benefits.  Already, there is some element of ``adaptive'' scheduling
due to human feedback (e.g., observers identify interesting targets to
be observed more frequently, time allocation committees decide how
much and when observing time will be made available).  Unfortunately,
quantifying these effects is extremely difficult.  One advantage of
following an algorithmic procedure is that any biases can be
recognized, simulated, and quantified.

As we demonstrated in Fig.~2, the use of adaptive
scheduling algorithm can significantly increase the number of planet
detections in a survey with a fixed amount of observing time.
Perhaps, more significantly, the additional planets found tend to have
the smallest velocity amplitudes and hence smaller masses of those
detected in the survey, as seen in Fig.~3.  For the survey parameters
which we considered, we found the least massive planet being detected
in a survey is a factor $\sim 2$ less massive when using our adaptive
scheduling algorithms than when scheduling observations randomly.  It
is important to appreciate that the increased the number of planet
detections does not require reducing the precision of measurements of
orbital parameters.  In fact, the same algorithms which increase the
number of planet detections simultaneously can improve the precision with
which most of orbital parameters are measured (Fig.~5).  While the precision is
comparable for planets with very large velocity amplitudes, for low
mass planets the adaptive algorithms typically measure orbital
parameters with more than an order of magnitude smaller uncertainties.
%


The adaptive algorithms presented in this paper address several
important challenges raised by previous studies.  In this paper, we
dramatically reduce the computational requirements for Bayesian
adaptive scheduling algorithms relative to Loredo (2004).  We
accomplish this by separating the integrals over orbital period from
the integrals over the other orbital parameters.  By assuming planets
on circular orbits, the remaining integrals become Gaussian integrals
which can be evaluated analytically.  The increased efficiency has
made several other advances possible.  The increased computational
efficiency of our algorithms allows us to conduct Monte Carlo
simulations of entire planet search programs and quantify the increase
in efficiency that adaptive scheduling algorithms offer.  More
significantly, our algorithms make it practical to perform a Bayesian
analysis of the orbital parameters even when there are extremely weak
constraints on the orbital parameters, such as when only a few
observations have been made.  This has allowed us to apply Bayesian
hierarchical modeling to simultaneously consider both possibilities
that a star has no planet and that the star has one planet,
extending previous Bayesian techniques which assumed there was a
single planet (Loredo 2004; Ford 2005, 2006).  By constructing hierarchical
models, adaptive scheduling algorithms naturally consider the
problems of planet detection and orbital parameter estimation
simultaneously.  We also present several generalizations of our
computationally efficient algorithms.  The generalized algorithms can
accommodate a variety of utility functions which can be customized to
the specific goals of an observational program.  These generalizations
also allow the adaptive scheduling algorithm to consider practical
complications such as stars of different brightnesses and observing
seasons.


This work also raises several new challenges.  Clearly, it would be
desirable to generalize our algorithms to reflect the full variety of
planetary systems.  For example, we assume that each star has a
maximum of one planet, while there are already $\sim$12 stars known to
have multiple planets.  In principle, it is easy to generalize our
hierarchical models to allow more multiple planets, but in practice
each additional planet would introduce an additional integral which
can not be evaluated analytically and dramatically increase the
required computations.

Additionally, we have assumed circular planetary orbits, while most
extrasolar planets have significant eccentricities.  We have conducted
some simulations in which we consider a population of stars with
planets on eccentric orbits but the adaptive scheduling algorithm
assumes circular orbits.  While the adaptive scheduling algorithms
still detect planets and measure orbital parameters more efficiently
than non-adaptive algorithms, the improvement in efficiency is less
than when applied to a population of stars with planets on circular
orbits.  Intuitively, we expect that planets on eccentric orbits could
benefit from adaptive scheduling algorithms even more than planets on
circular orbits.  Therefore, we expect that the reduction in
efficiency is due to the adaptive scheduling algorithm not having
access to the appropriate model.  Again, in theory, it is
straightforward to include eccentric orbits in a Bayesian analysis,
but this would introduce an additional two integrals which can not be
evaluated analytically and hence would significantly increase the
require computations.  Incorporating eccentric orbits into adaptive
scheduling algorithm could be accomplished by brute force, or it might
be sufficient to use approximate models which expand the orbital
motion in the eccentricity.  

Our simulations have also assumed that planetary perturbations are the
only cause of variations in the star's radial velocity.  In practice,
many stars appear to have intrinsic variability commonly known as
stellar ``jitter'' which can be comparable to or exceed the
observational uncertainties in the radial velocity measurements (Saar,
Butler, \& Marcy 1998).  We have conducted some simulations in which
we consider a population of stars where each star has a Gaussian
jitter, but the adaptive scheduling algorithm assumes no jitter.  The
primary effect of the jitter is to increase the false alarm rate
(fraction of stars with no planets for which the Bayesian analysis
determines the probability of having no planet is less than 0.1\%).
While the threshold for announcing a planet detection can be altered
to maintain a 0.1\% false alarm rate, such a treatment is simplistic
and does not properly account for the unknown jitter.  The problems
posed by jitter can be mitigated by replacing the observational
uncertainties with the observational uncertainties added in quadrature
to the amount of jitter expected based on the star's spectral
properties.  Still, this approach could result in poor performance
when the estimate for the stellar jitter is inaccurate.  Indeed, one
of the advantages of Bayesian analysis is that it can naturally allow
for noise sources of unknown magnitude.  A more rigorous analysis
would treat the stellar jitter of each star as an unknown along with
the orbital parameters.  Unfortunately, adding the stellar jitter as a
model parameter introduces an additional integral and requires
additional computation time.  For the purpose of testing adaptive
scheduling algorithms when confronted with jitter, it may be useful to
consider a special case in which the observational uncertainties are
the same for each observation.  In this case, the integral over the
unknown stellar jitter can be performed separately from the other
integrals and reduces to a sum of exponential integrals which 
can be performed analytically (Cumming 2004).

This paper has focused its attention on adaptive scheduling algorithms
for radial velocity planet searches.  Targeted astrometric planet
searches such as those planned for the Space Interferometry Mission
(SIM) are another obvious application of our methods.  Given the 
cost and finite lifetime of space missions such as SIM, observing time
is extremely valuable and it is even more important to make the best
use of the available observing time.  Successfully incorporating Bayesian
adaptive design would require that the observation schedule not be fixed
far in advance by logistical constraints.  Mission designers should aim
to allow for frequent upload of revised target lists.
In principle, the two types of
planet searches are quite similar, with the main difference that
astrometric surveys can measure the stellar position in two dimensions
while radial velocity surveys can measure the stellar velocity in only
one dimension.  Therefore, we expect that adaptive scheduling
algorithms could also provide a significant improvement in the
efficiency of the SIM planet searches, resulting in more planets being
discovered, planet masses and orbital parameters being determined more
accurately, and {\em significantly increasing the sensitivity of SIM
to nearly-Earth-mass planets and multiple planet systems}.  For this
potential benefit to be realized the SIM design must be sufficiently
flexible that the observing schedule and target stars can be chosen
with a lead time much less than the duration of the mission,
preferably a lead time of a month or less.  To fully simulate such
an adaptive scheduling algorithm, it will be important to incorporate
the practical observing constraints and costs (e.g., possible pointing
directions limited, time require to slew to different positions).

\acknowledgments

We thank Debra Fischer, Jeremy Goodman, Geoff Marcy, Scott
Tremaine, and an anonymous referee for their suggestions.
This research was supported in part by the Miller Institute for Basic
Research, NASA grants NAG5-10456 and NNG04H44g, and by NASA through
Hubble Fellowship grant HST-HF-01195.01A awarded by the Space
Telescope Science Institute, which is operated by the Association of
Universities for Research in Astronomy, Inc., for NASA, under contract
NAS 5-26555.

\newpage

\begin{figure}
\label{FigN2k}
\plotone{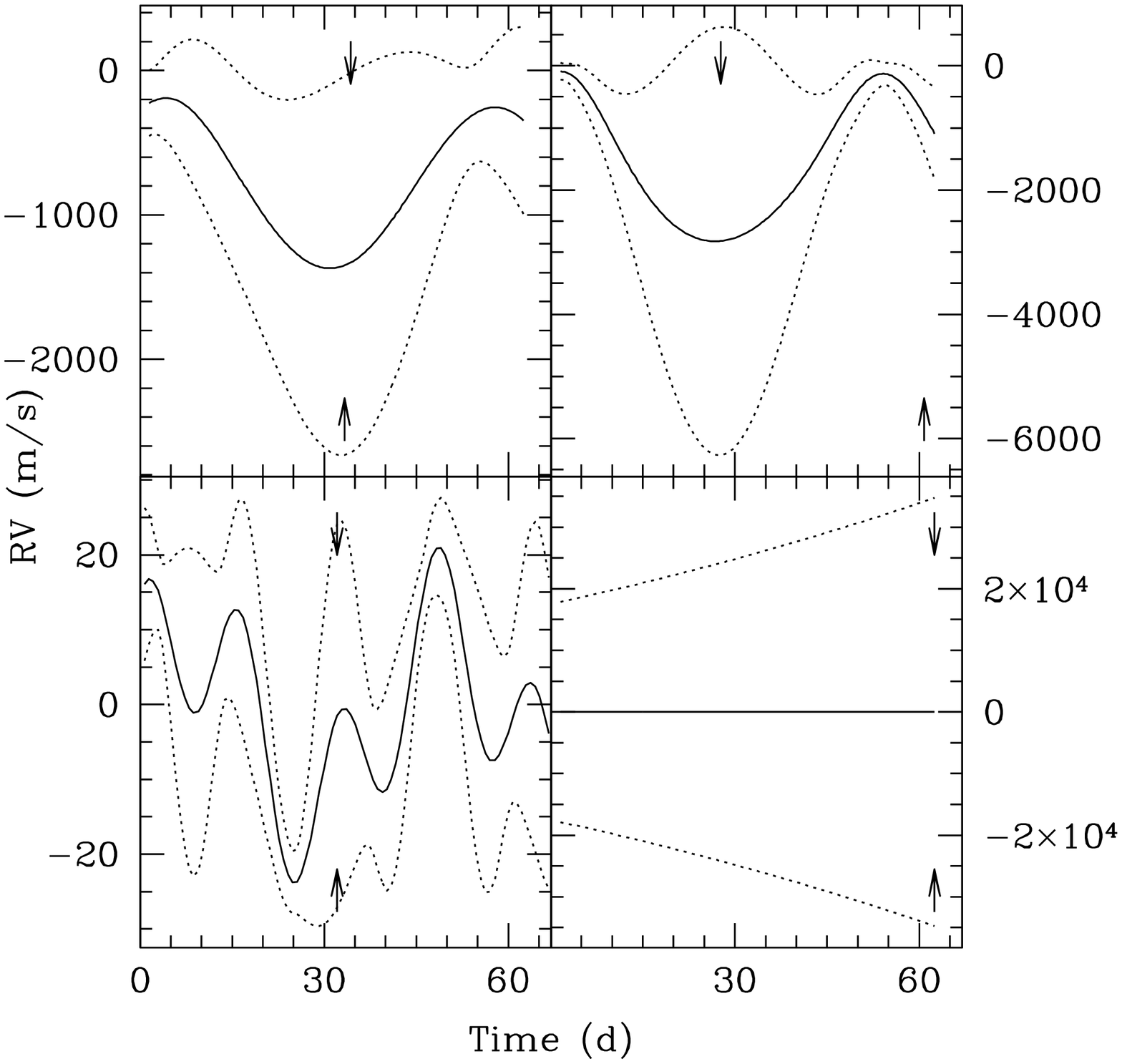}
\caption{Here we show the expected value for the radial velocity
(solid line) during October and November 2004 (tick marks very five
days) of four different target stars from the N2K project.  The dotted
lines show the 95\% confidence intervals for $p(v|d)$ as a function of
time.  These predictive observations were based on observations taken
between January 10 and July 11, 2004 (not shown) as a part of the N2K
survey.  The arrows near the bottom of each panel indicate the time
when the entropy of the predictive distribution is maximized.  The
arrows near the bottom of each panel indicate the time when the
standard deviation of the predictive distribution is maximized.  The
predictive distributions are based on 4 (top left), 5 (top right), 11
(bottom left), and 6 (bottom right) observations.  The best-fit
orbital periods are near 53d (top left), 52d (top right), 16 d (bottom
left), and 27d (bottom right).
%
}
\end{figure}

\begin{figure}
\label{FigHd88133}
\plotone{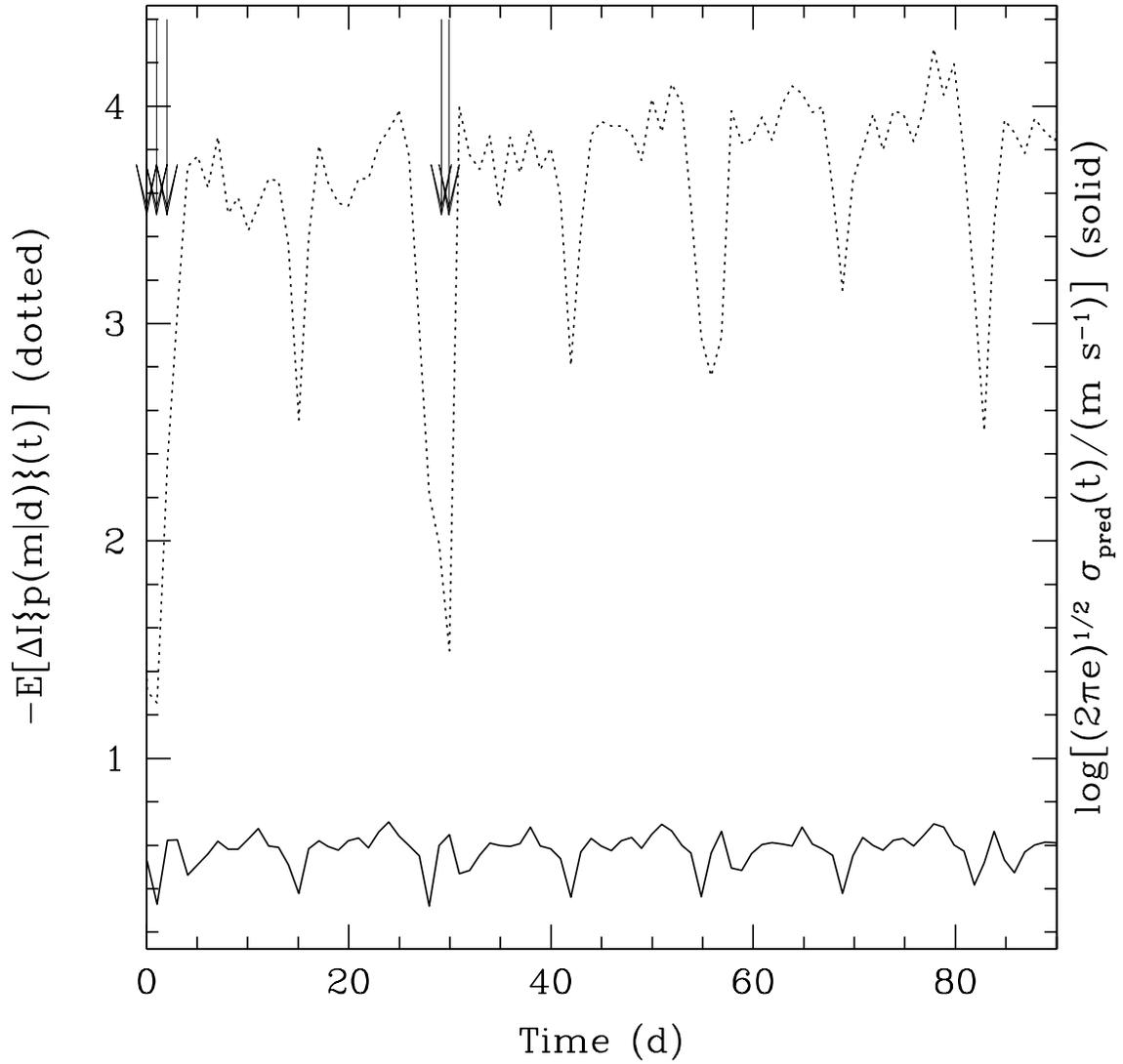}
\caption{Here we show the expected change in the expectation of the information
content following a single observation as a function of the time, $t$ (dotted line).  
This is to be compared to the log of the standard deviation of the predictive distribution (solid line).  
The figure is based on the first five of the observations of HD 88133, the first planet discovered by the N2K project (Fischer et al. 2004).  The times of the observations are indicated with vertical arrows.  While the two curves show similarities, the most/least favorable times according to an information theory analysis are not always coincident with what would be predicted using only the variance of the predictive distribution.
}
\end{figure}

\begin{figure}
\plotone{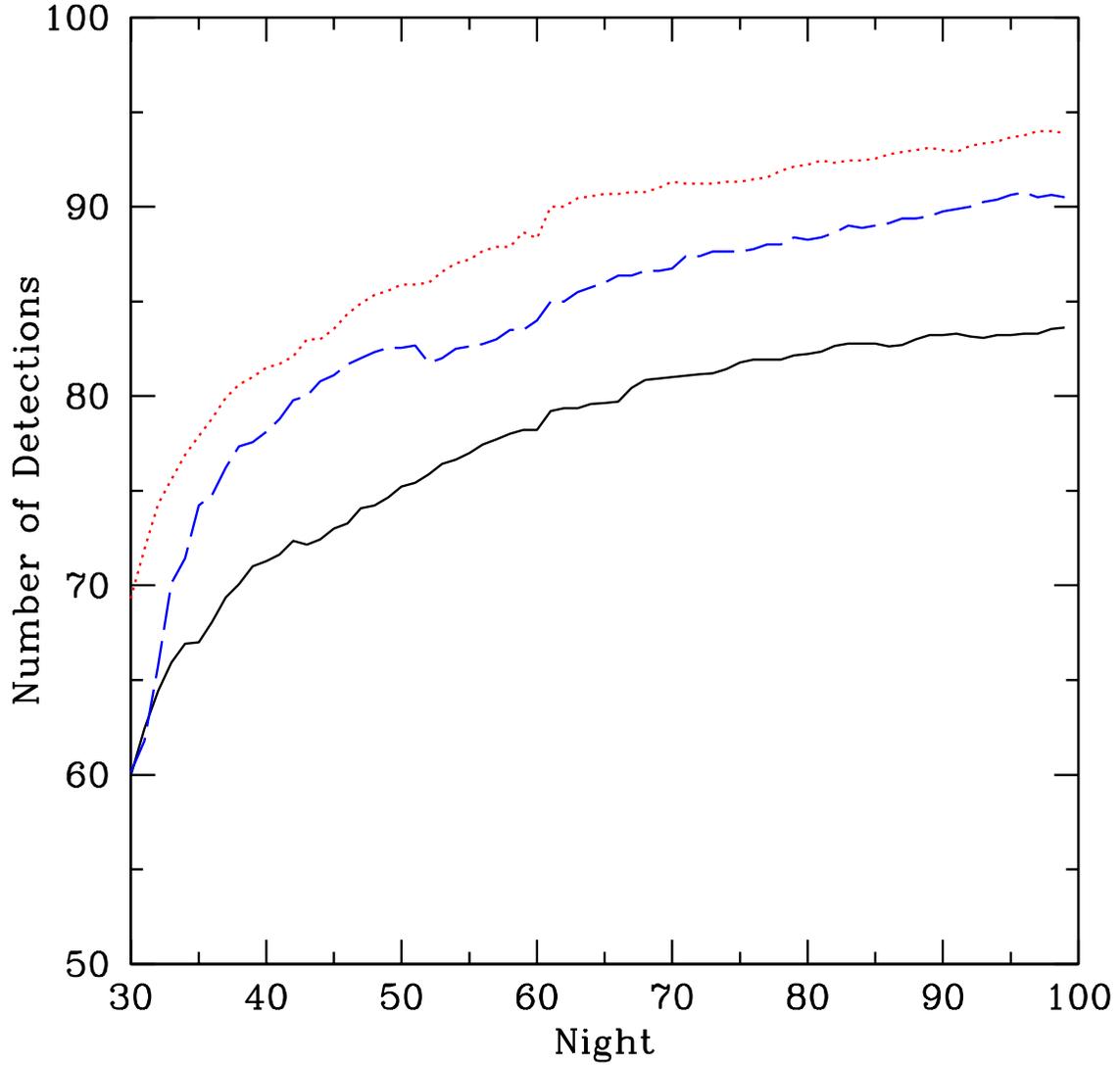}
\caption{This figure shows the number of planets detected (stars for
which the posterior probability of the no-planet model,
$p(0|\vec{d})$, is less than 0.1\%) as a function of the number of
observing nights.  The different line styles present the results for
different scheduling algorithms: regular (solid black), the adaptive
algorithm from \S\ref{SecMultiMaxEntropy} (dashed blue), and the 
%
%
adaptive algorithm from \S\ref{SecMultiAltMargin} (dotted red), which
%
%
maximizes the detections at the expense of accuracy in orbital
parameters by marginalizing over orbital parameters (compare to Fig.\
5).  The results for each scheduling algorithm have been averaged over
ten simulated observing programs.
}
\end{figure}

\begin{figure}
\plotone{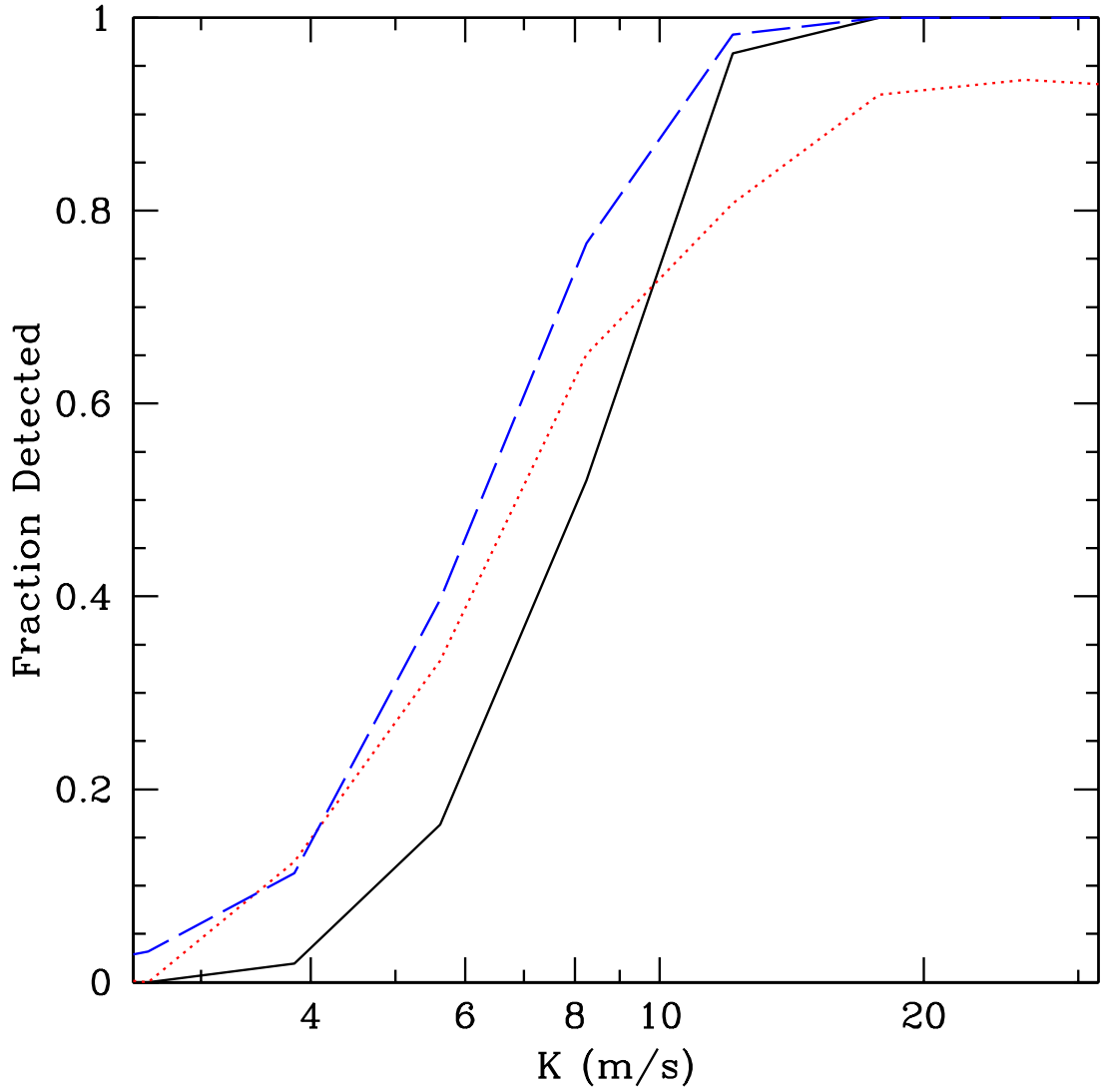}
\caption{Here we show the fraction of planets detected as a function
of the velocity semi-amplitude, $K$.  In these simulations each
observation had an observational uncertainty of $\sigma_{\mathrm obs}
= 3$m/s.  The planet searches based on both adaptive algorithms are
significantly more efficient at detecting planets with $K\sim 1-3
\sigma_{\mathrm obs}$ than the fixed schedule (solid black line).  The
line styles are as in Fig.~2.  }
\end{figure}

\begin{figure}
\plotone{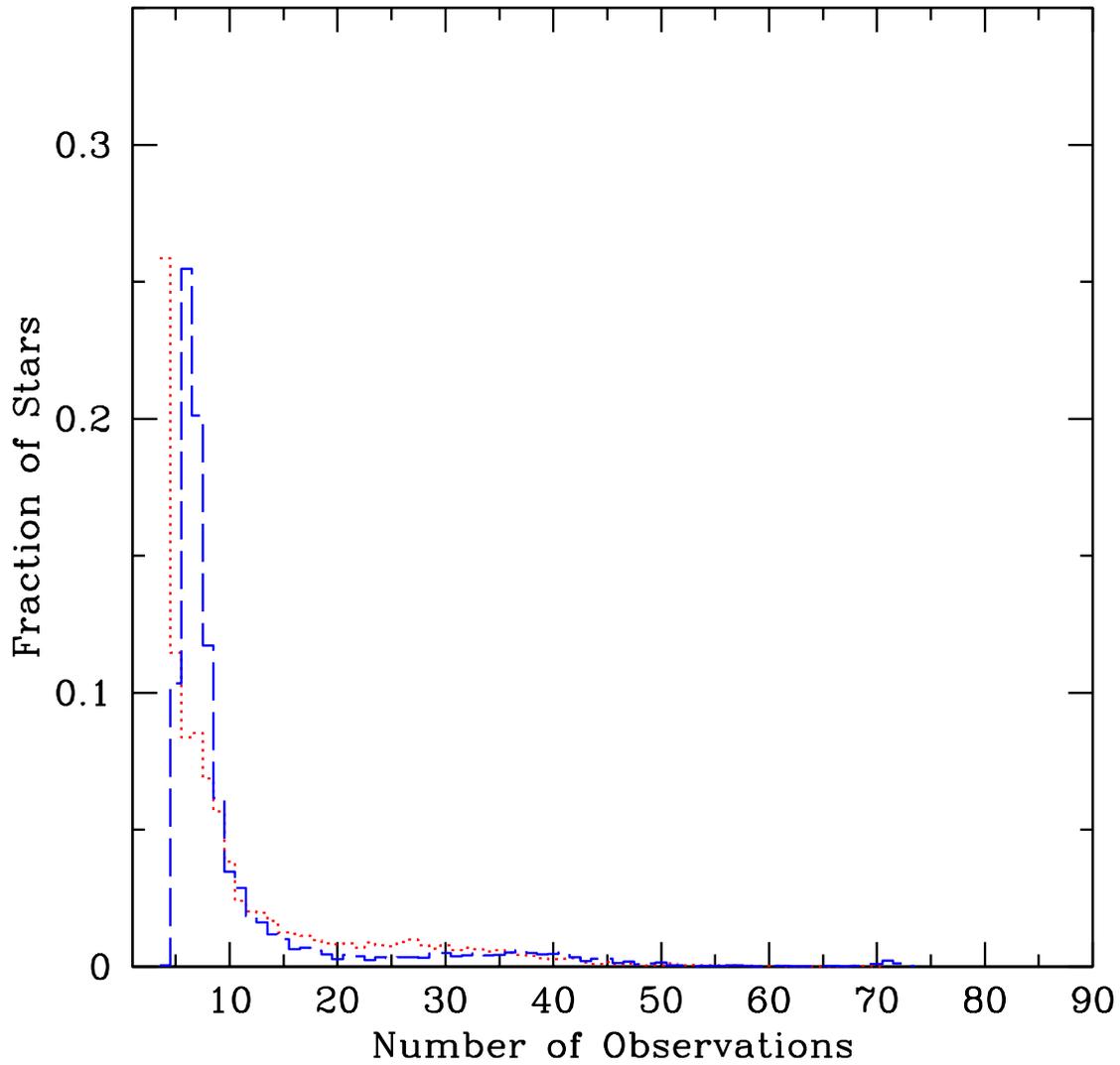}
\caption{In this figure we present a histogram of the number of time a star was
observed.  For the regular scheduling algorithm, each star was
observed ten times (not shown).  The adaptive algorithm from
\S\ref{SecMultiMaxEntropy} (dashed blue) and the adaptive algorithm from
%
%
\S\ref{SecMultiAltMargin} (dotted red) both devote a large number of
%
%
observations to a few stars, significantly increasing the sensitivity
to low-mass planets around these stars.  }
\end{figure}

\begin{figure}
\plotone{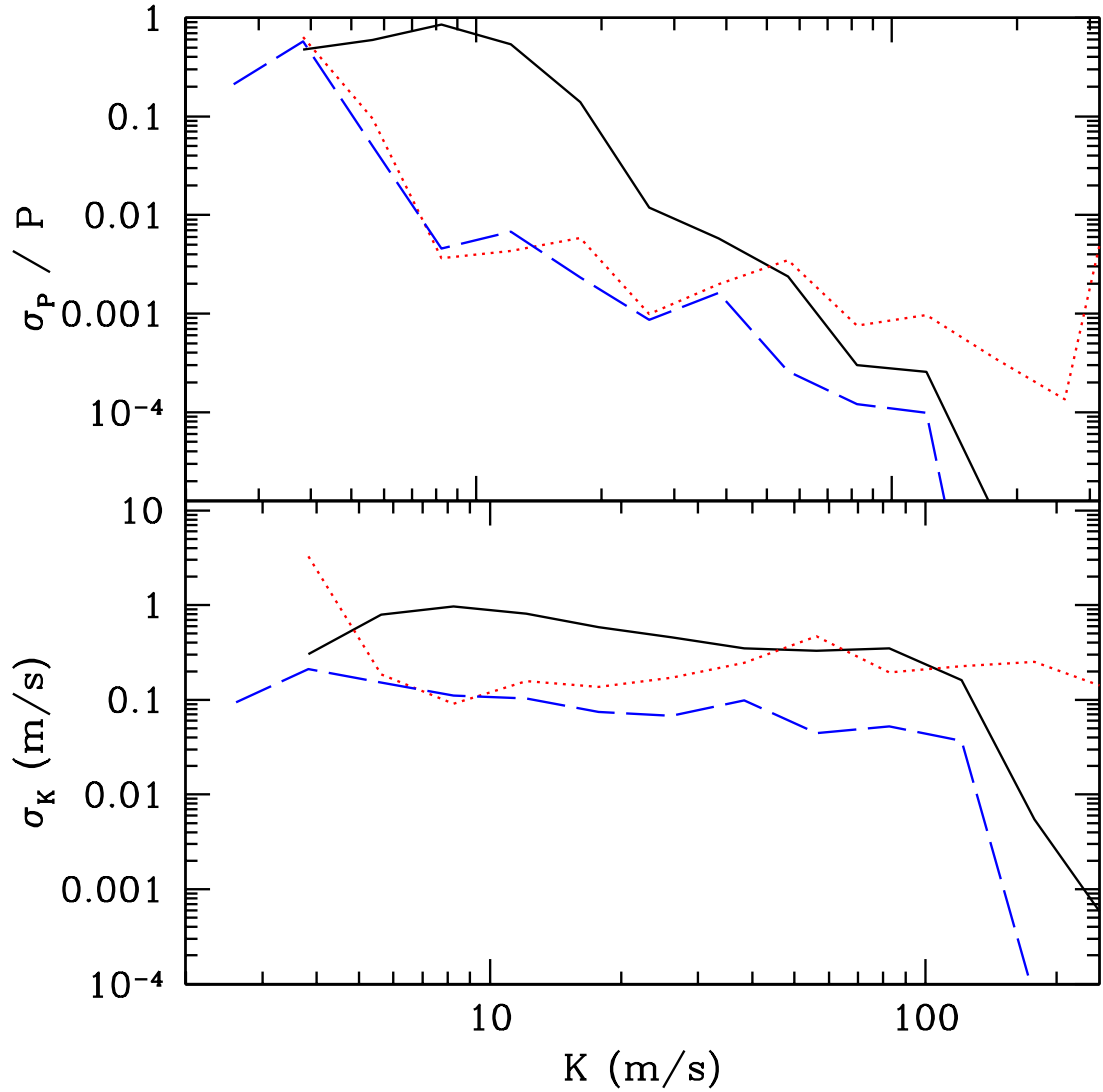}
\caption{Here we show the median precision of the measurements of the orbital parameters, $P$ (top panel) and $K$ (bottom panel), as a function of the velocity semi-amplitude, $K$.  Both adaptive scheduling algorithms typically perform significantly better than the fixed schedule (solid black line).  However, for large velocity amplitudes, the adaptive algorithm presented in \S\ref{SecMultiMaxEntropy} often allocates a only small number of observations to planets with large velocity amplitudes and hence the orbit determinations are not as precise as with the fixed schedule.  The line styles are as in Fig.~2.
}
\end{figure}

\end{document}